# Harnessing the power of complex light propagation in multimode fibers for spatially resolved sensing

**DARCY L. SMITH**[1,2,3,4,†,*], **LINH V. NGUYEN**[3,4,†], **MOHAMMAD I. REJA**[1,3], **ERIK P. SCHARTNER**[1], **HEIKE EBENDORFF-HEIDEPRIEM**[1], **DAVID J. OTTAWAY**[1,2] AND **STEPHEN C. WARREN-SMITH**[3,4]

[1]*Institute for Photonics and Advanced Sensing and The School of Physics, Chemistry and Earth Sciences, The University of Adelaide, Adelaide, SA 5005, Australia*
[2]*Australian Research Council Centre of Excellence for Gravitational Wave Discovery (OzGrav)*
[3]*Future Industries Institute, University of South Australia, Mawson Lakes, SA 5095, Australia*
[4]*Laser Physics and Photonics Devices Laboratory, University of South Australia, Mawson Lakes, SA 5095, Australia*
*\*darcy.smith@adelaide.edu.au*
[†]*These authors contributed equally to this work*

**Abstract:** The propagation of coherent light in multimode optical fibers results in a speckled output that is both complex and sensitive to environmental effects. These properties can be a powerful tool for sensing, as small perturbations lead to significant changes in the output of the fiber. However, the mechanism to encode spatially resolved sensing information into the speckle pattern and the ability to extract this information is thus far unclear. In this paper, we demonstrate that spatially dependent mode coupling is crucial to achieving spatially resolved measurements. We leverage machine learning to quantitatively extract this spatially resolved sensing information from three fiber types with dramatically different characteristics and demonstrate that the fiber with the highest degree of spatially dependent mode coupling provides the greatest accuracy.

## 1. Introduction

The propagation of light through complex media results in behavior that has traditionally been seen as a hindrance to optical systems which should be avoided or compensated for. Precision instruments, such as gravitational-wave detectors, require ultrahigh vacuums to avoid material interactions [1]. When propagation through media cannot be avoided, such as ground-based astronomy, complex adaptive optics are used to correct for the distortions [2]. Sometimes, light propagation through complex media can also be used to advantage. Random lasers are a striking example where lasing is achieved without a defined resonator [3, 4] to produce high brightness but low coherence light for speckle-free imaging [5].

Waveguides can lead to complex light propagation even in the absence of material inhomogeneity via their discrete transverse modes whose phase relationship varies with propagation. For an optical fiber with a fixed configuration, the resulting output field is inherently deterministic but highly complex in practice due to interference between the many guided modes, interaction with the environment, and mode coupling. Recently, there has been significant interest in understanding and exploiting the transmission of light through multi-mode optical fiber (MMF), due to their inherent ability to transport a greater amount of information than single mode fiber (SMF) [6]. A field that has particularly leveraged the power of MMF is imaging, with applications such as endoscopy [7, 8] and fluorescence imaging [9]. MMFs have also been utilized for applications such as high-resolution spectrometry [10, 11], laser pulse characterization [12], fiber amplifier beam shaping [13], and control of nonlinear light generation [14]. A major limitation of MMFs in imaging applications is their vulnerability to mode coupling and changes in modal phase relations induced by external perturbations. Efforts to overcome this include obtaining a basis set of perturbation insensitive modes [15] or training neural networks to extract desired information from the noisy output [16, 17]. Such sensitivity can also be used to advantage, for example, tailoring a desired output through deliberately applied perturbations to an MMF [18].

Optical fiber sensing can be considered as the inverse procedure to imaging. Rather than keeping the fiber and the surrounding environment static and mapping from the input of the fiber to the output, the optical input is fixed, while a map is built between the fiber's output and the changes to the light propagation through it due to external perturbations. Possible perturbations include temperature, strain, pressure, or even biochemical interactions, all of which can induce optical path length changes in the optical fiber [19]. When coherent light is coupled into an MMF, these environmental perturbations lead to complex changes in the transverse spatial intensity output (specklegram) [20]. Typically, the specklegram is analyzed by comparing the measured output to a reference image using correlation techniques [21-23] or machine learning [24, 25]. We have recently shown that the combination of deep learning with MMF interferometry allows for accurate sensing even in the presence of significant environmental noise [26]. In these cases, the sensor integrates the measurand along the length of the fiber, effectively yielding a single longitudinal measurement point.

The true power of optical fibers for sensing is multi-point or distributed sensing [27-29]. This is generally achieved in single mode fiber (SMF) using time or frequency domain interrogation methods, such as Brillouin scattering based distributed fiber sensing [30, 31]. The following question then arises – is it possible to perform spatially resolved sensing directly using the sensitive, but complex, MMF interference? Recent reports indicate this is indeed possible, with deep learning being applied to extract spatial information from an MMF's output for sensing [32-34]. Such demonstrations have shown the ability to classify the location of a perturbation but were restricted to qualitative analysis. Nevertheless, important insights can be drawn from such work, like that deep neural network (DNN) accuracy increases for sensing locations along the fiber which are closer to the light interrogation apparatus [33], and that a ring core fiber supporting several weakly-coupled mode groups could achieve greater accuracy with less training data compared to conventional graded-index multi-mode fiber [34].

In this paper, we demonstrate that mode coupling in an MMF is essential for spatially resolving perturbations in the resulting multimode interferometric output. We demonstrate this concept by comparing the performance of three distinctly different MMF architectures for the task of spatially resolved temperature sensing: a graded-index fiber, a microstructured optical fiber, and a sapphire crystal optical fiber. While quantitative and distributed sensing can be achieved for all three fibers, the sapphire fiber displays better longitudinal sensing resolution capabilities due to significant diameter inhomogeneities formed by the crystal fabrication process. This allows spatial information to be more rigorously encoded in the complex interferometric output through mode coupling. This approach opens a new pathway for distributed fiber sensing directly with the raw, complex interference pattern output of MMFs, significantly simplifying hardware requirements and broadening the range of fiber materials that can be deployed.

## 2. Concept

*Mode coupling theory for spatially resolved fiber sensing*

In this paper, we investigate both theoretically and experimentally the ability to perform multi-point fiber sensing directly with the multimode interference in an MMF. We show through the use of an MMF mode coupling matrix theory framework [35], which considers multimode light propagation, mode coupling and fiber gain/loss, that mode coupling in an MMF gives the resulting output longitudinal spatial resolution along the fiber axis. It is concluded that a perfectly longitudinally symmetric fiber which is free from mode coupling has no ability to longitudinally resolve perturbations, and that this symmetry must be broken in order to resolve the position of such perturbations. This concept is shown graphically in Fig. 1.

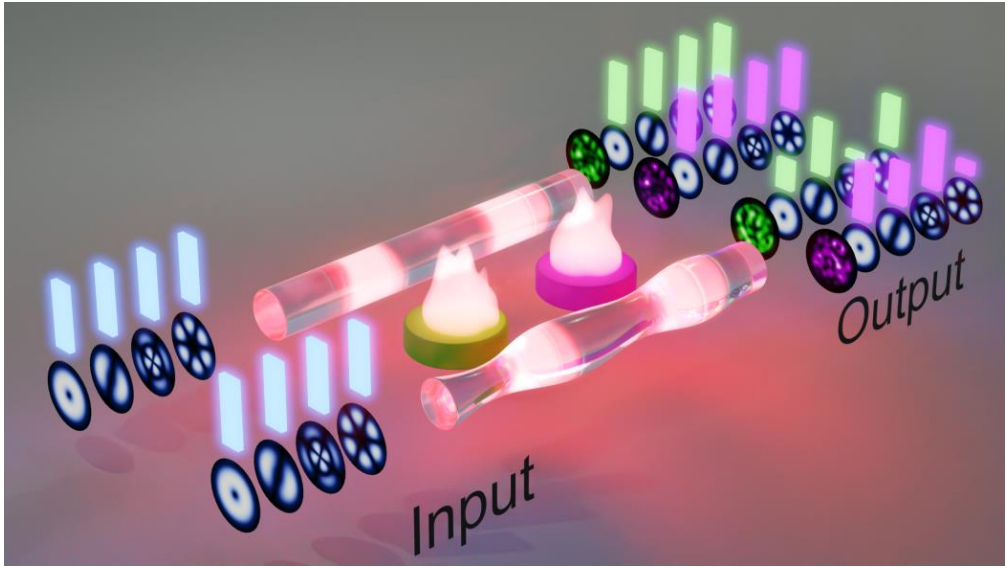

Fig. 1. Concept of spatially resolved sensing enabled by distributed mode mixing: comparison between a longitudinally invariant fiber and a fiber with diameter variations. The light within a multimode optical fiber consists of a superposition of eigenmodes, each carrying a portion of the total power propagating in the fiber. A perturbation on the fiber will induce mode-dependent phase changes, which will manifest as a change in the output of the fiber, whether this is the spatial interference pattern (specklegram) shown in the figure, or the wavelength domain interference spectrum used in this work. In this figure, two different longitudinal positions that experience identical perturbations are shown in green and purple, resulting in the corresponding color-coded modal amplitudes. For a perfectly longitudinally invariant fiber (left), the position of this perturbation will be indistinguishable at the output, as the modes will travel through the fiber free from coupling and power redistribution, thus rendering the effect of mode-dependent optical path length changes position-independent. In the case of an optical fiber with longitudinal variations (right), mode coupling leads to modal power redistribution, rendering the effect of these path length changes position-dependent, hence allowing for longitudinally resolved sensing.

In this framework, the electric field propagating through an MMF with $N$ orthogonal transverse eigenmodes can be expressed as:

$$\mathbf{E}(x, y, z) = \sum_{j=1}^{N} A_j(z) \hat{\mathbf{e}}_j(x, y),$$

(1)

where $\hat{\mathbf{e}}_j$ denotes the unit electric field profile of the $j^{\text{th}}$ orthonormal transverse eigenmode, and $A_j$ contains the amplitude and phase of this mode. Assuming fixed eigenmodes, the light can be described fully by the vector $\mathbf{A}$,

$$\mathbf{A} = \big(A_1(z), \dots, A_N(z)\big).$$

(2)

Propagation of this field through an optical system can be expressed as:

$$\mathbf{A}^{(\text{out})} = M^{(t)} \mathbf{A}^{(\text{in})},$$

(3)

where $M^{(t)}$ is an $N \times N$ matrix, termed the 'propagation matrix' of the system, which includes the effects of phase accumulation, mode dependent gain/loss, and mode coupling.

The propagation of the light through a multimode fiber can be modelled by dividing the fiber into $K$ adjacent sections, the $k^{\text{th}}$ section having length $L^{(k)}$. The $k^{\text{th}}$ propagation matrix can be decomposed as:

$$M^{(k)} = V^{(k)} \Lambda^{(k)} U^{(k)*},$$

(4)

where $\Lambda^{(k)}$ represents coupling-free propagation, given by

$$\Lambda^{(k)} = \begin{bmatrix} \exp\left(\frac{1}{2} g_1^{(k)} - i\phi_1^{(k)}\right) & & 0 \\ & \ddots & \\ 0 & & \exp\left(\frac{1}{2} g_N^{(k)} - i\phi_N^{(k)}\right) \end{bmatrix},$$

(5)

and $U^{(k)}$ and $V^{(k)}$ denote the input and output coupling matrices to and from the $k^{\text{th}}$ section respectively. The real part of the exponential arguments in Eq. (5) describe mode-dependent loss or gain, where $g_j^{(k)}$ is the gain coefficient of the $j^{\text{th}}$ mode in the $k^{\text{th}}$ section of fiber. The imaginary part, $i\phi_j^{(k)}$, where $\phi_j^{(k)} = \beta_j L^{(k)}$ and $\beta_j$ denotes the propagation constant of the $j^{\text{th}}$ mode, describes the accumulated phase of the $j^{\text{th}}$ mode as it passes through the $k^{\text{th}}$ section of fiber. This is mode-dependent due to effects such as modal dispersion and mode-dependent chromatic dispersion.

Propagation of light through the entire fiber can be represented by a cascade of $K$ propagation matrices as in Eq. (6).

$$M^{(t)} = M^{(K)} M^{(K-1)} \ldots M^{(k)} \ldots M^{(2)} M^{(1)}.$$

(6)

Now consider that the $k^{\text{th}}$ section of fiber experiences a perturbation, inducing mode dependent optical path length changes and hence phase changes in this section. Such a perturbation can be modelled by a diagonal perturbation matrix $\tilde{\Lambda}$ [Eq. (7)], which describes the change in optical path length that each mode experiences:

$$\tilde{\Lambda} = \begin{bmatrix} \exp(-i\Delta\phi_1) & & 0 \\ & \ddots & \\ 0 & & \exp(-i\Delta\phi_N) \end{bmatrix}.$$

(7)

Hence, the total propagation matrix in the presence of this perturbation is given by Eq. (8).

$$M^{(t)} = M^{(K)} M^{(K-1)} \ldots V^{(k)} \tilde{\Lambda} \Lambda^{(k)} U^{(k)*} \ldots M^{(2)} M^{(1)}.$$

(8)

The matrix multiplication $U^{(k+1)*} V^{(k)}$, represents projection of the field from the modal basis in the $k^{\text{th}}$ section to the $(k+1)^{\text{th}}$ section, i.e., mode coupling. If the modal bases are identical, this term reduces to the identity matrix, and no mode coupling occurs. Otherwise, off-diagonal terms will be present.

Consider now a longitudinally invariant MMF which experiences no mode coupling. $M^{(t)}$ will be the product of diagonal uncoupled propagation matrices and products of input and output coupling matrices projecting the fiber's modal basis onto itself, each time producing the identity matrix. Hence, $M^{(t)}$ will itself be diagonal, with the $j^{\text{th}}$ diagonal element representing the sum of the gain/loss and phase accumulated by the $j^{\text{th}}$ mode. In the presence of a perturbation, $\tilde{\Lambda}$, the $j^{\text{th}}$ phase change will simply be added to the $j^{\text{th}}$ diagonal element of $M^{(t)}$, regardless of which section of fiber the perturbation was applied to. Hence, such a longitudinally symmetric fiber possesses no mechanism by which the perturbation can be spatially resolved.

Now consider an MMF with longitudinal variance, and hence coupling between the modal bases of each adjacent section of the fiber. $M^{(t)}$ will no longer be diagonal, as off-diagonal terms representing the coupling of power between modes will be present. As such, it will also be dependent on the order of multiplication of the constituent propagation matrices. Hence in the presence of optical path length changes due to a perturbation, the fiber's output will be dependent on the position of this perturbation along the fiber. Furthermore, the degree to which the fiber's output is sensitive to this spatial information will be dependent on the degree to which adjacent propagation matrices commute, suggesting that stronger mode coupling will more rigorously encode spatially resolved information in the fiber's output.

It should be noted that even in the case of strong mode coupling, the transmission matrix can still be diagonal if a particular basis is used, namely the spatial eigenmode basis of the said mode-coupling MMF [36]. If the input field has the spatial profile of one of these spatial eigenmodes, then it will be reconstructed at the output of the fiber. However, these modes are not truly propagation invariant in the sense that they do not propagate through the MMF with a fixed form while scaling in phase and gain/loss, as is the case for the transverse eigenmode basis discussed above. These spatial eigenmodes are linear combinations of transverse eigenmodes, which still couple between each other in the presence of fiber inhomogeneities as they propagate.

Any perturbation or length change in this case will distort the total transmission matrix, and thus the defined spatial eigenmode basis no longer holds. Since the physical mode-coupling process between transverse eigenmodes is not dependent on the basis chosen, perturbations at different locations along the MMF will lead to different outputs. It is only for the case of truly propagation-invariant transverse eigenmodes which are free from coupling throughout propagation, that is, in a longitudinally uniform MMF, that sensing information along the fiber will not be encoded.

The information encoded by mode coupling is, however, included in the output field in a complex and non-trivial way. We have previously shown that deep learning proves an effective method for extracting single-point sensing information from a complex MMF output [26]. In this work, deep learning is used as a tool for extracting distributed sensing information from this output.

*Generation of a wavelength intensity spectrum from a specklegram*

In our experiment (Sec. 3), we convert the specklegram to a wavelength domain interference spectrum by splicing the MMF to SMF. This is done for practicality as the spliced fiber connection is stable and compatible with high temperatures. To understand that this approach is equivalent to measuring the specklegram, consider again that Eq. (6) represents the propagation of light through an MMF. Coupling from this MMF into an SMF, and the subsequent propagation of the light, can be represented by multiplying the right-hand side of Eq. (6) on the left by $V^{(\text{SMF})} \Lambda^{(\text{SMF})} U^{(\text{SMF})*}$, where $\Lambda^{(\text{SMF})}$ is a $1 \times 1$ matrix simply describing the change in amplitude and phase of the light propagating through the SMF. If $N$ eigenmodes are propagating in the $K^{\text{th}}$ (final) section of the MMF, then $U^{(\text{SMF})*} V^{(K)}$ will be a $1 \times N$ matrix representing the projection of these $N$ modes onto the single mode basis. This projection, and consequently the power propagating in this single mode, is dependent on the modal power distribution and relative phases of the $N$ modes, which in turn is dependent on the wavelength of the propagating light. By measuring the intensity output by the SMF at various wavelengths, one obtains a wavelength intensity interference spectrum. Effectively, the information carried by the MMF transmission is projected onto a single value when coupled into the SMF, but one can restore this information by scanning across a range of wavelengths, thus transforming the information from the modal domain to the wavelength domain. Consequently, if the multimode speckle contains spatially resolved information as a result of mode coupling, then the wavelength spectrum, which is dependent upon the speckle, will also have this information encoded.

### 3. Experimental method

*Experimental overview*

Two experiments were performed to demonstrate the spatially resolved sensing concept described in this paper, (1) a localized heating experiment and (2) a distributed sensing experiment. In both experiments we tested the effect of spatial heat distributions on the output of three different optical fiber sensors; a sapphire optical fiber (75 µm diameter, Micromaterials), an in-house fabricated suspended-core fiber (SCF) (10 µm core diameter) [37], and a graded-index (GRIN) fiber (OM1, FS). An overview of the method used to collect wavelength interference spectra from these fibers is shown in Fig. 2(a), with a detailed schematic of the sensors displayed in Fig. 2(b). An SMF-MMF-SMF configuration was formed from a single SMF-MMF splice and the back-reflection from the end facet of the sensing fibers, converting the MMF speckle information into an interference spectrum in the wavelength domain.

The wavelength interference spectra were collected from the fibers using a Hyperion si155 swept source interrogator (Luna Technologies). These spectra existed in the interrogator as 20 000 wavelength points across the 1460-1620 nm range, but were decimated down to 1 000 points for computer memory purposes. Low-frequency Fourier filtering was performed to remove a strong back-reflection from the sensing fiber/SMF splice in the sapphire fiber spectra, and performed on the spectra from the other two fibers for consistency.

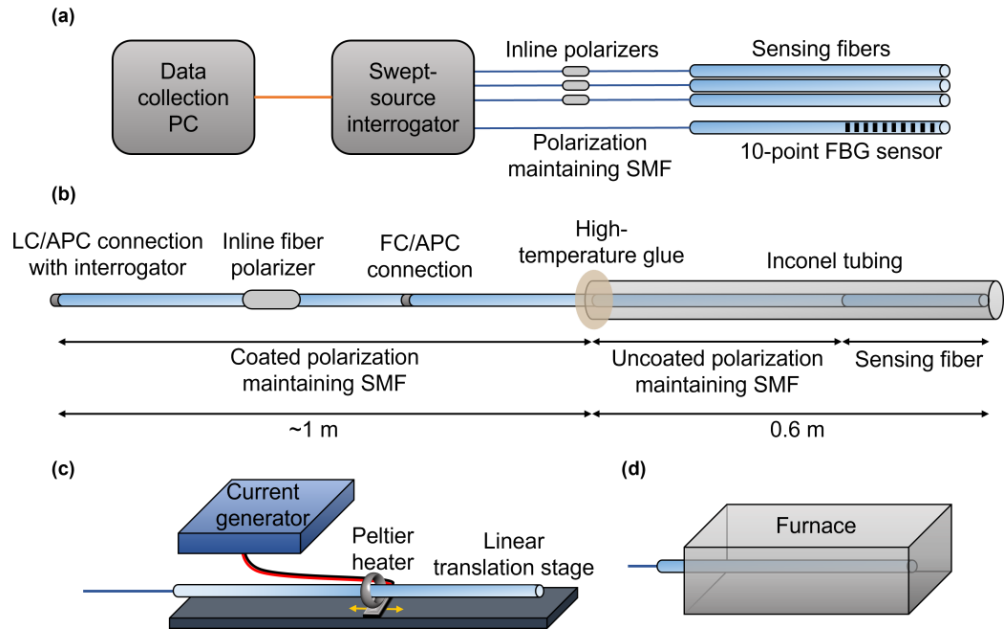

Fig. 2. Experimental setup. (a) Overview of the setup used to collect wavelength intensity spectra from the fibers under different temperature distributions. A swept-source interrogator recorded the spectra produced by the reflected light coupled back into SMF from the cleaved end facets of the MMF sensing fibers. Polarization maintaining SMF was used in conjunction with inline polarizers so as to prevent polarization mode coupling within the SMF. (b) Schematic and dimensions of the sensors. The length of sensing fiber for each of the sensors was 0.13 m for the sapphire fiber, 0.24 m for the suspended-core fiber and 0.29 m for the graded-index fiber. (c) The experimental setup for the localized heating experiment. The fibers were subjected to a local

heating perturbation through means of an aluminum block attached to a Peltier heater, translated via a linear translation stage. (d) The experimental setup for the furnace heating experiment. The fibers were subjected to various temperature distributions by being placed in a furnace, with the distributions being varied by manually translating the fibers relative to the furnace and applying a range of furnace temperatures per sensor position.

In the localized heating experiment, represented in Fig. 2(c), a localized and constant heat distribution was translated along the fiber length. The aim of this experiment was to compare the degree to which the location of a perturbation along the fiber leads to differences in the optical output. To assess this effect, the correlation of the MMF output to a reference spectrum was measured [see Eq. (9)] to determine the degree to which the output is dependent on the position of the applied perturbation.

The second experiment, represented in Fig. 2(d), subjected the fibers to a range of different temperature distributions generated by a furnace, and used deep learning to extract the spatially resolved sensing information from their outputs. The aim of this experiment was to test and compare the three fibers using the temperature distribution of the furnace as an example of a spatially resolved sensing task. The ability for a deep learning model to converge to an accurate model based on the spectra collected from each of the fibers was indicative of the degree to which useful, spatially resolved sensing information was encoded in their outputs and could be quantitatively extracted.

*Fiber sensors*

Table 1. A comparison between the three sensing fibers for a few relevant properties, including cross-sectional image/refractive index profile, the numerical apertures, core diameters and an approximate number of supported modes.

|  | Sapphire | Suspended-core | Graded-index |
|---|---|---|---|
| Cross section or refractive index profile | 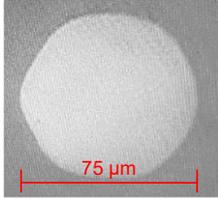 | 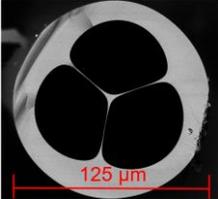 | 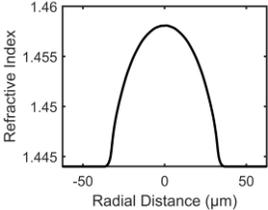 |
| Numerical aperture | 1.46 | 0.2 | 0.275 |
| Core diameter | 75 μm | 7 μm | 62.5 μm |
| Approximate number of supported modes | 100 000 | 200 | 200 |

The technical characteristics of the three fibers are compared in Table 1, with a summary of their characteristics as follows. The sapphire fiber is fabricated via crystal growth, rather than by preform fabrication as in the manufacturing process of conventional glass optical fibers. As such, the sapphire fiber has significant longitudinal diameter variations the order of ±10 μm [~25% variation, see Fig. 3], leading to strong spatially dependent mode coupling as light propagates through it. The SCF is designed for temperature sensing, supporting the propagation of modes with relatively large differences in thermo-optic response compared to solid fibers

[38]. In contrast to the sapphire fiber, both the SCF and GRIN fibers are designed and fabricated with highly consistent diameter (< 3% variation).

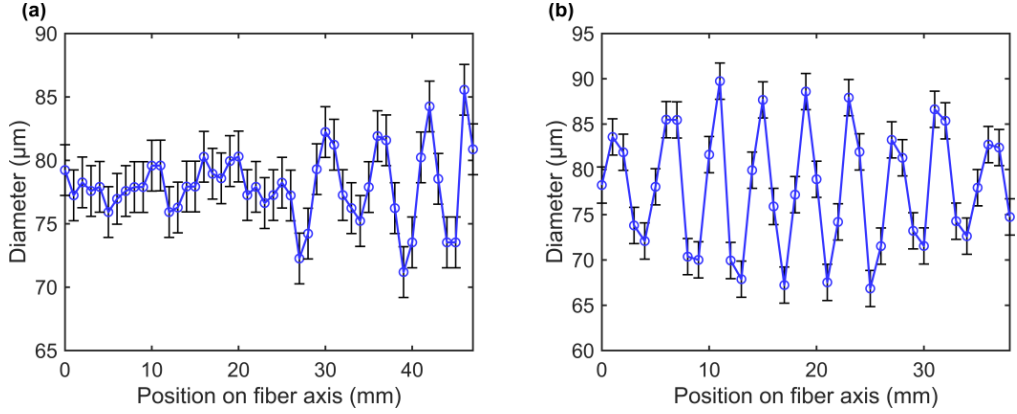

Fig. 3. Variation in sapphire fiber diameter as a function of position. (a) The diameter of an arbitrarily selected ~50 mm length of sapphire fiber as a function of position along the fiber axis. (b) The diameter of a 40 mm length of sapphire fiber, chosen for its visibly strong variation in diameter, as a function of the position along the fiber axis. The lengths of fiber which generated the plots in (a) and (b) were selected from the same, originally longer length of sapphire fiber, as used in the main experiment.

A 10-point FBG sensor was placed alongside the three sensing fibers to characterize the temperature distributions that the fibers were subjected to and provide temperature labels for the deep learning. This FBG sensor consisted of ten femtosecond laser ablation written gratings inscribed in a suspended-core microstructured optical fiber [39-41]. The ten FBGs were spaced 15 mm apart, thus performing temperature sensing over a 135 mm length.

*Zero-normalized cross-correlation*

For the first, localized heating experiment, the changes experienced by each fiber's wavelength intensity spectra $I(\lambda)$ were quantified using the zero-normalized cross correlation function [ZNCC, Eq. (9)] of each spectrum against a reference spectrum. The ZNCC is defined as

$$Z[I(\lambda)] = \frac{\sum_\lambda (I_0 - \bar{I}_0)(I - \bar{I})}{[\sum_\lambda (I_0 - \bar{I}_0)^2 \sum_\lambda (I - \bar{I})^2]^{\frac{1}{2}}},$$

(9)

where $I_0(\lambda)$ denotes the reference intensity spectrum and the bar denotes the average intensity of a given spectra.

*Deep learning*

Two neural networks were employed for the second, distributed sensing experiment: a multi-layer perceptron model [42] with four hidden layers and 672 842 trainable parameters, and a linear regressor with 9 760 trainable parameters. The linear regressor learns an output which is simply a linear combination of the input values. In-depth summaries of the neural networks and the training process are given in Appendix A.

The two different DNN architectures were chosen to assess the degree to which spatially resolved sensing information was encoded in the wavelength spectra of the fiber outputs, by correlating this with the convergence ability and accuracy of the trained models. That is, the aim of this experiment was not necessarily to produce an accurate temperature sensor, but rather to assess the ability of the fibers in projecting sensing information from the spatial domain to the modal domain and finally the wavelength domain. Hence a weaker model was used to induce greater contrast in its training and performance when trained on outputs from the three fibers, allowing for clearer conclusions to be made.

## 4. Results

### *Localized heating: correlation analysis*

The first set of results are from the localized heating experiment. Wavelength spectra were recorded while the heater was sequentially moved to five equidistant positions along the length of the fibers. The correlation coefficient for each spectrum was calculated against a reference spectrum from the first temperature profile, and the average correlation coefficient value [Eq. (9)] for each position was calculated and are displayed in Fig. 4(a). The thermal distribution at each heating position is shown in Fig. 4(b). A lower correlation coefficient for a spectrum indicates relatively larger decorrelation from the reference spectrum. Hence unique correlation values for different heating locations, continuously decreasing from the reference spectrum, indicates longitudinal resolution in the fiber output.

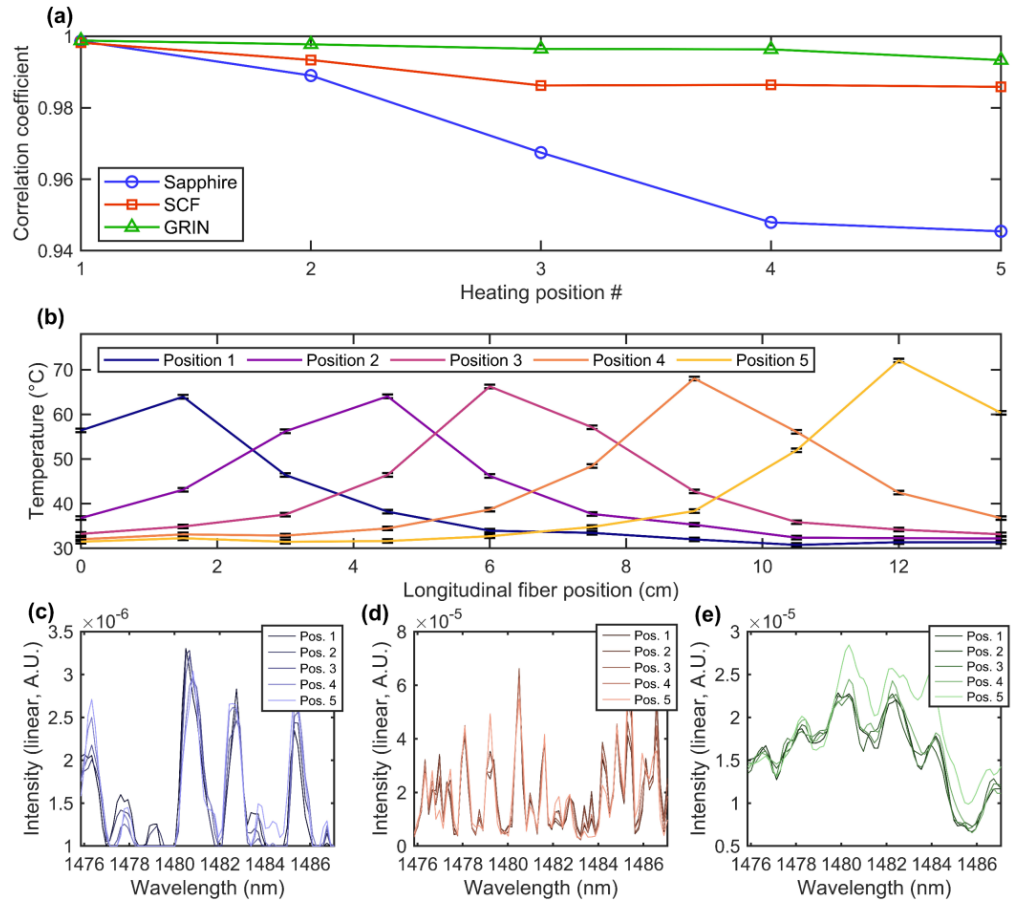

Fig. 4: Results from the localized heating experiment. (a) The average correlation coefficient of spectra collected during each 30 min heating position against the reference spectrum (the first spectrum collected from heating position #1). (b) Heat distributions from the localized heating experiment. Shown are five heat distributions from the five heating positions across the 135 mm of sensing length which the fibers were subjected to. These temperature measurements were taken using the FBG sensor. (c)-(e) Example spectra over a zoomed in wavelength range from all five heating positions for the sapphire, suspended-core, and graded-index fiber, respectively.

Fig. 4(c)-(e) show example wavelength spectra for the sapphire, suspended-core, and graded-index fibers, respectively, for the five heater locations. The spectra are zoomed in to a ~10 nm range, to display the spectral features more clearly.

The results indicate that the recorded spectra from each fiber have different dependencies on the position of the thermal perturbation. It is clearly shown that the spectrum of the sapphire fiber is more dependent on the position of the heater, as indicated by the correlation coefficient continuously decreasing with heater translation. The spectra collected from the SCF and GRIN fiber do not display this dependence as clearly, with the respective correlation coefficients staying closer to unity over the whole range of the fiber lengths. As argued in the theory, weak mode coupling implies that the effect of a perturbation on the fiber's output will only be weakly dependent on the position of the perturbation, leading to similar outputs for the different temperature profiles. In contrast, strong location dependency indicates strong mode coupling, hence the correlation coefficient decreases quicker as the location of the perturbation is shifted.

With reference to the specific fibers used, it is the significant longitudinal variance of the sapphire fiber, shown in Fig. 3, which induces mode coupling continuously along the fiber axis and direction of light propagation, hence enhancing the spatial resolution of sensing information in the multimode interferometric output. The SCF and GRIN fibers are expected to exhibit much less mode coupling due to being drawn with minimal longitudinal variations. We note that the change in correlation for the SCF is greater than the GRIN, which is attributed to the higher overall temperature sensitivity of the SCF and thus sensitivity to subtle changes in the temperature distribution as the heater is moved.

*Distributed heating: deep learning analysis*

The results from the second experiment show the temperature predictions made by the DNNs trained on wavelength spectra from each fiber under different temperature distributions produced by a furnace. To subject the sensing fibers to various temperature distributions, two dimensions of variation were applied. First, data was collected with the fibers in seven different locations relative to the furnace. This was achieved by manually translating the fibers along their fiber axis relative to the furnace. Three of these positions are shown in Fig. 5, with Fig. 5(a) showing distributions from a position where the temperature sensing points were all fully within the furnace, hence a uniform distribution, Fig. 5(c) showing a sensor position where the sensor was located outside of the furnace, hence only the end of the fibers receiving a small amount of heat, and Fig. 5(b) showing a sensor position between these extremes. For each sensor position, the furnace was left to cool from 500°C to 50°C over a period of 20 h.

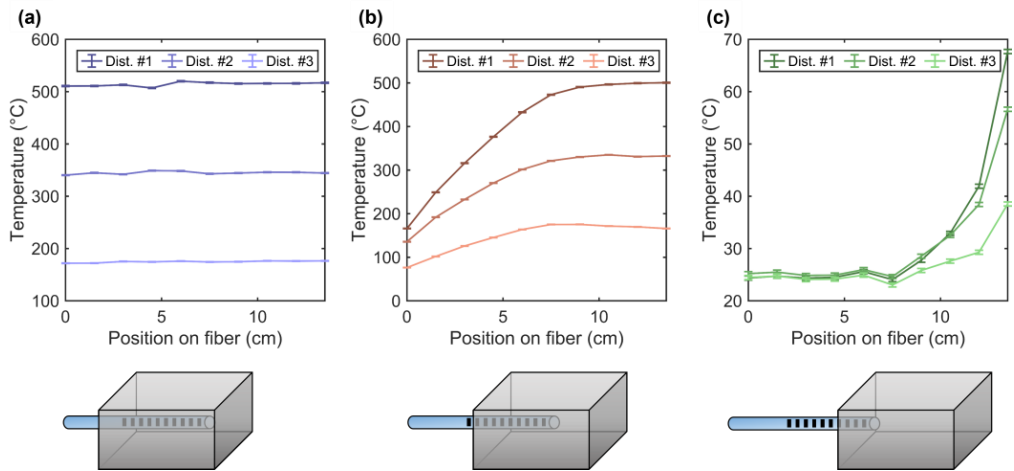

Fig. 5. Heat distributions from the furnace heating experiment. (a)-(c) Temperature distributions from three different sensor positions. Seven sensor positions were used to collect spectra from the sensing fibers while the furnace was set to cool from 500°C to room temperature over a 20 hr period. Within each plot/sensor position, the temperature distributions from three different furnace temperatures are shown. Shown are the positions of the sensors relative to the furnace.

Fig. 6(a)-(c) show the predictions made by the linear regressor, while Fig. 6(d)-(f) shows those made by the MLP. The associated RMS error for each set of predictions is displayed. The temperature labels were collected by the 10-point FBG sensor. The DNNs had 10 nodes in their output layer, hence trained to predict 10-point temperature distributions. The full set of predictions on all 10 sensing points (for the 47 600 test spectra) are included in each plot of Fig. 6(a-f), while the error as a function of each sensing point is shown in Fig. 6(g-h).

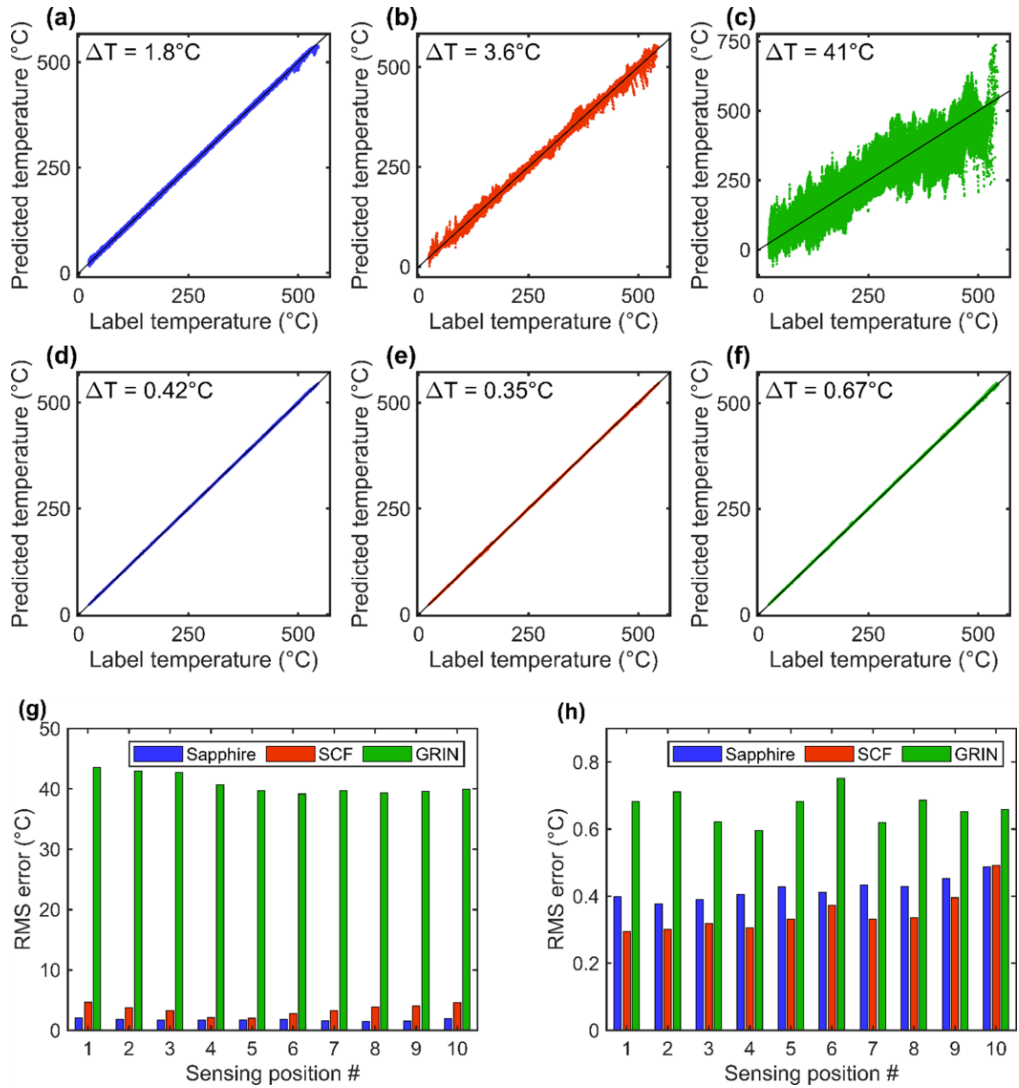

Fig. 6: Predictions made by the DNNs on the 10-point temperature distributions from the wavelength interference spectra. Predictions were made on a test set of spectra, which were kept aside during all preliminary model training, and only used to produce the figures seen in this work. (a)-(c) Predictions made by the linear regressor. (d)-(f) Predictions made by the MLP. (a) and (d) show the predictions from DNNs trained on sapphire spectra, (b) and (e) SCF spectra, and (c) and (f) graded-index (GRIN) fiber spectra. (g) and (h) Model accuracy for each model and each sensing position. RMS error for predictions made by neural networks on the test set of spectra for the three fibers, by sensing position. (g) Accuracies per sensing position for the linear regressor. (h) Accuracies per sensing position for the multilayer perceptron.

The multi-layer perceptron (MLP) is able to predict the 10-point temperature distribution under which the unseen test set of spectra were collected with high accuracy for all three fibers. Given the heavily reduced number of trainable parameters compared to the MLP, the linear regressor model overall converges to a higher loss value yet displays a greater contrast in this convergence between the models trained on the three different fibers' outputs.

As discussed in Sec. 2, mode coupling should be required to perform spatially resolved fiber sensing with an otherwise unaltered fiber. The result we have obtained that the MLP performs accurately when trained on spectra from either of the three fibers suggests that the level of inherent mode coupling present in the two otherwise longitudinally invariant fibers is sufficient to encode spatially resolved sensing information in the spectra, at least to the 15 mm resolution (the distance between sensing points) measured in this experiment. This may be induced by the sensing parameter itself, as temperature variations lead to refractive index modulations.

The results from the linear regressor show more clearly a quantitative difference in DNN performance. The model achieves an RMS error of 1.8°C when trained on spectra from the sapphire fiber, 3.6°C for the SCF spectra, and 41°C for the GRIN fiber spectra. The lower-capacity linear model used gives a greater contrast in model performance, allowing conclusions to be made regarding the ability for the model to extract longitudinally resolved sensing information from the spectra from the different models. In agreement with our theoretical framework and the correlation analysis results, the sapphire fiber shows better sensitivity to the position of a perturbation, or in the more general case, the spatial distribution of a perturbation.

## 5. Discussion and conclusion

The results presented support the concept that deep learning can be used to extract distributed sensing information directly from the output of an MMF. The presence of mode coupling facilitates the encoding of this information in the output of the fiber, leading to spatial resolution along the direction of light propagation which may be exploited for sensing.

The localized heating experiment demonstrated that the output of the sapphire fiber in the presence of a localized perturbation is more positionally dependent than that of the SCF and GRIN fibers. This is reflected in Fig. 4(a), as the sapphire spectrum decorrelates from the reference spectrum in the presence of translation of the heat perturbation in a clear and continuous manner which is not seen in the spectra of the SCF and GRIN fibers.

We then applied deep learning to decode spatially resolved sensing information from the output of an MMF, comparing the convergence ability and accuracy of the models when trained on spectra from the three fibers. The overall accuracy of the higher-capacity MLP can be attributed to the small amount of unavoidable mode coupling that these fibers possess, as well as the nature of the temperature distributions achieved by the furnace, and how these were far from arbitrary. As such, sufficient distributed temperature information is encoded in the spectra of all three fibers for the DNN to learn.

The performance of the linear regressor, however, gives more insight into how clearly each fiber encodes distributed sensing information in their outputs. This low-capacity model is able to extract this information from the sapphire fiber spectra to an RMS error of 1.8°C, half that of the SCF, and 20 times less than that of the GRIN fiber. This is aided by the mode coupling which occurs along the sapphire fiber's length, encoding longitudinally resolved sensing information in the multimode interferometric output, which can be more efficiently extracted through deep learning techniques.

Our work has clearly demonstrated that mode coupling is the essential feature that allows spatially resolved information to be encoded in the interferometric output of a multimode fiber. While strong mode coupling from large waveguide inhomogeneities can lead to enhanced distributed sensing capability, we observed that quantitative distributed sensing can still be achieved in relatively longitudinally invariant fibers. This is likely due to inherent manufacturing tolerances and the variations induced by the sensor parameter itself (thermo-optic effect in our example). Our method can thus be generally applied to distributed fiber sensing even using commercially available fiber, provided a suitable method of extracting the distributed sensing information, such as deep learning, can be applied. Further, we believe these findings will have wider implications in optical sensing, such as the use of other complex media.

**Appendix A: Deep neural networks**

Two neural network architectures were used for taking MMF wavelength intensity spectra as their inputs and mapping to 10-point temperature distributions. The first was a multilayer perceptron (MLP) with four hidden layers and ReLU nonlinear activation [43], while the second was a simple linear regressor. The input layer for both was the 975-point normalized wavelength intensity spectra and the output layer was the 10 temperature predictions.

**Table A1. Architecture of the multilayer perceptron. The size and number of trainable parameters of each layer are given.**

| Layer type | Output shape | Number of parameters |
|---|---|---|
| Input layer | (1,975) | - |
| Dense hidden layer | (1,512) | 499 712 |
| Dense hidden layer | (1,256) | 131 328 |
| Dense hidden layer | (1,128) | 32 896 |
| Dense hidden layer | (1,64) | 8 256 |
| Output layer | (1,10) | 650 |
| Total number of trainable parameters | | 672 842 |

**Table A2. Architecture of the linear regressor. The size and number of trainable parameters of each layer are given.**

| Layer type | Output shape | Number of parameters |
|---|---|---|
| Input layer | (1,975) | - |
| Output layer | (1,10) | 9 760 |
| Total number of trainable parameters | | 9 760 |

The neural networks were built using the Keras modular deep learning library for Python. Summaries of the two architectures are displayed in Table A1 (MLP) and Table A2 (linear regressor). Training was performed on an NVIDIA GeForce RTX 2070 GPU. The 150 epoch-long training process took around two hours for the 4-layer model, and around 20 min for the linear regressor.

The complete dataset consisted of 476 000 spectra for each fiber. Following normalization as per an L2 normalization technique and preceding training, this set was shuffled and split into three groups: 80% for training, 10% for validation, and 10% for testing. The training set was used to train the model and the validation set was used to test the prediction ability and monitor the performance of the model during preliminary training runs. The test set was kept aside and only used to make predictions using the final trained model and generate the results seen in this work.

The deep learning model began the training process with randomized parameters. Then, mini-batches of 64 spectra were randomly selected from the training set and passed through the model. The loss (cost function) of the model was calculated using the mean-squared-error (MSE) metric, and backpropagation performed to adjust the model towards one with a lower loss [44]. This process was repeated with the mini-batches of 64 training samples, until all samples had been used. This concluded one epoch of training. Training was performed for 150 epochs, with the end-of-epoch model exhibiting the lowest loss kept as the final model.

Backpropagation and stochastic gradient descent was performed using the Adam optimizer [45], with a learning rate of $10^{-2}$ for the linear regressor, and $10^{-3}$ for the MLP. These values were chosen through brief experimentation as suitably sensitive values for the given architectures.

The training and validation loss as a function of epoch is shown in Fig. 7. In none of the cases is underfitting (characterised by lack of convergence) or overfitting (characterised by the simultaneous convergence of the training loss and divergence of the validation loss) present [46], indicating a suitable selection of architectures and hyperparameters for the given dataset.

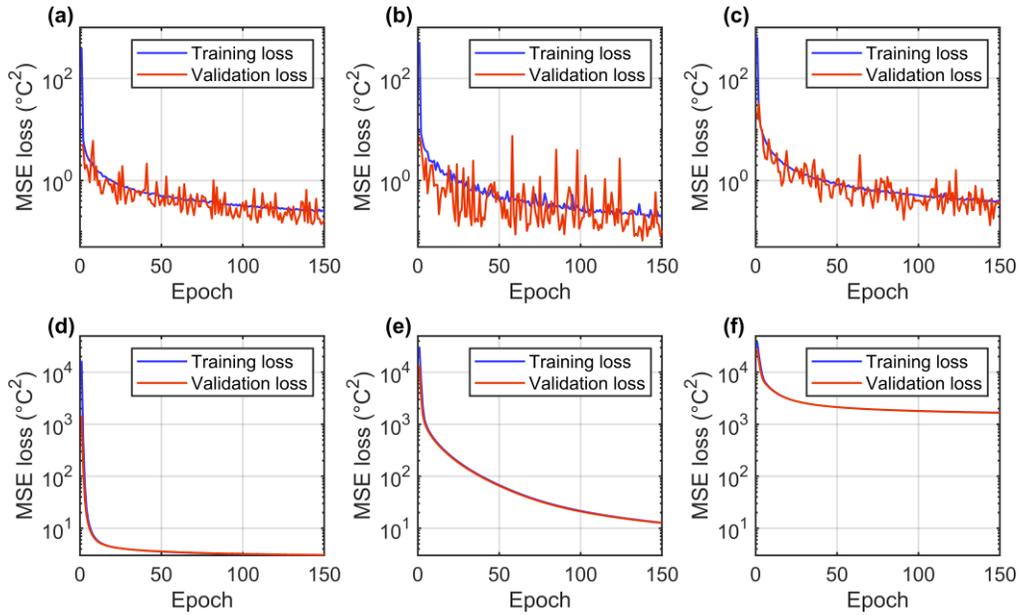

Fig. 7. Deep neural network loss history. (a)-(c) Training (blue) and validation (red) mean-squared error (MSE) loss as a function of epoch for the MLP trained on the (a) sapphire fiber spectra, (b) SCF spectra and (c) graded index (GRIN) fiber spectra. (d)-(f) Training and validation mean-squared error loss as a function of epoch for the linear regressor trained on the (d) sapphire fiber spectra, (e) SCF spectra and (f) GRIN spectra.

**Funding.** Stephen Warren-Smith is supported by an Australian Research Council (ARC) Future Fellowship (FT200100154). Darcy Smith and Md. Istiaque Reja are supported by Australian Government Research Training Program Scholarships. This work is supported by the Australian Research Council (ARC) under CE170100004. Fiber fabrication was performed at the OptoFab node of the Australian National Fabrication Facility utilizing Commonwealth and SA State Government funding.

**Acknowledgements.** The authors thank Alastair Dowler, Evan Johnson, and Minh Hoa Huynh from the University of Adelaide for their contribution in fabricating the multimode suspended-core fiber used in this work.

**Disclosures.** S.C.W.S. is a director of HT Sensing Pty. Ltd., a company that manufactures optical fiber sensors. HT Sensing Pty. Ltd. did not contribute to or participate in this research in any way.

**Data availability.** Data underlying the results presented in this paper are not publicly available at this time but may be obtained from the authors upon reasonable request.